\documentclass[twocolumn]{article}

\usepackage{geometry}
\usepackage{graphicx}
\usepackage{float}
\usepackage{subfig}
\usepackage{multicol}
\usepackage{multirow} %

\usepackage{url}

\usepackage{textcomp} %

\usepackage{booktabs} %

\makeatletter
\newcommand{\gobblepars}{\@ifnextchar{\par}{\expandafter\gobblepars\@gobble}{}}
\makeatother
\newcommand{\minisec}[1]{\smallskip\noindent\textbf{{#1}.}\ \ \gobblepars}

\usepackage[depythontex]{pythontex/pythontex}

\begin{pycode}
import sys
from make_tables import *
from make_graphs import *
attr_header = r" & \multicolumn{2}{c}{\multirow{2}{*}{Self-reported}} & \multicolumn{4}{c}{Google inferred on}\\ \cmidrule(l){4-7}  &  &  & \multicolumn{2}{c}{Services} &   \multicolumn{2}{c}{Web} \\ & in & out & in & out & in & out"
prc_header = r" & \multicolumn{2}{c}{Sex services} & \multicolumn{2}{c}{Sex web} & \multicolumn{2}{c}{Age services} & \multicolumn{2}{c}{Age web}\\ & in & out & in & out & in & out & in & out"
\end{pycode}

\title{The Accuracy of the Demographic Inferences\\ Shown on Google's Ad Settings\thanks{A version of this paper will appear at ACM WPES 2018.  A version has already appeared as ICSI tech.\@ report number TR-16-003 in 2016.}}

\newcommand{\email}{\url}

\author{Michael Carl Tschantz\thanks{International Computer Science Institute}\\\email{mct@icsi.berkeley.edu}
\and
Serge Egelman\footnotemark[2]\ \thanks{University of California, Berkeley}\\\email{egelman@icsi.berkeley.edu}
\and
Jaeyoung Choi\footnotemark[2]\ \thanks{Delft University of Technology}\\\email{jaeyoung@icsi.berkeley.edu}
\and
Nicholas Weaver\footnotemark[2]\ \footnotemark[3]\\\email{nweaver@icsi.berkeley.edu}
\and
Gerald Friedland\footnotemark[2]\ \footnotemark[3]\\\email{fractor@icsi.berkeley.edu}}

\begin{document}

\def\etal{{\it et al.~}}

\newenvironment{packed_enum}{
\begin{enumerate}
  \setlength{\itemsep}{1pt}
  \setlength{\parskip}{0pt}
  \setlength{\parsep}{0pt}
}{\end{enumerate}}

\newenvironment{packed_item}{
\begin{itemize}
  \setlength{\itemsep}{1pt}
  \setlength{\parskip}{0pt}
  \setlength{\parsep}{0pt}
}{\end{itemize}}

\maketitle

\begin{abstract}
Google's Ad Settings shows the gender and age that Google has inferred about a web user.  
We compare the inferred values to the self-reported values of 501 survey participants.  We find that Google often does not show an inference, but when it does, it is typically correct.  %
We explore which usage characteristics, such as using privacy enhancing technologies, are associated with Google's accuracy, but found no significant results.
\end{abstract}

\section{Introduction}

\begin{figure*}[!t]
\begin{center}
\includegraphics[width=0.65\textwidth]{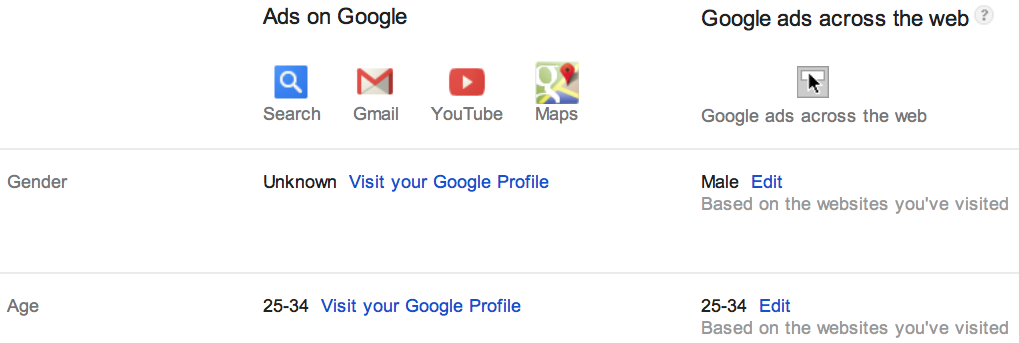}%
\end{center}
\caption{Screen shot of part of Google's Ad Settings webpage while logged in with a Google account using Safari. \textcopyright Google}
\label{fig:settings-small}
\end{figure*}

Google's Ad Settings offers users a window into the model that Google learns about them from online tracking and their account settings~\cite{google}.
Users may see inferences Google made about them at\\
\centerline{\url{https://www.google.com/settings/ads}}\\
Figure~\ref{fig:settings-small} shows a screen shot of the first author's settings from 2016, when we conducted this study (the page has since changed). %
The page provides two predictions each for the user's gender and age: one based
upon the information Google uses for its web services and one based upon
the information Google uses as part of its web-wide ad network.  
The page allows editing the inferences.
Google provides some information about how it works~\cite{google},
but
questions
remain about the accuracy of Google's profiles.

To study the accuracy of Google's predictions and how they are associated with user behaviors, attitudes, and usage of privacy enhancing technologies (PETs), 
we conducted a survey. 
We asked participants for their ages, genders, computer usage habits, attitudes, PETs usage, and for a copy-and-paste of the content of their Ad Settings page.
We compared their supplied demographics to the age range and gender provided by Google
and examined how various factors are associated with accuracy.

We find that Google's predictions tend to be accurate when Google makes them, but that Google often makes no inference.
In particular, Google rarely makes predictions for logged out users.
While we document that Google's accuracy for some subgroups of users is far from its overall accuracy, we do not find statistical significance for any such association after adjusting for the large number of hypotheses examined in this exploratory work.

We believe we are the first look at the accuracy of Google's inferences on Ad Settings with a survey.
We provide a new point of reference for understanding Google's ability to infer attributes of users.
Additionally, we believe this paper is a reasonable starting point for larger-scale confirmation studies.
We make additional information available in the appendix and data available at\\
{\small\mbox{}\hfill\mbox{}\verb|http://www.icsi.berkeley.edu/~mct/pubs/wpes18/|\mbox{}\hfill\mbox{}}

\section{Related Work}

Numerous studies have looked at how Google track users (e.g., \cite{guha10imc,balebako12w2sp,wills12wpes,liu13hotnets,barford14www,lecuyer14usenix,datta15pets,tschantz15csf,englehardt14man,nath15mobisys,book15arxiv}).
Datta et al.\ experimented on Ad Settings to determine how they impact the ads shown and how browsing behaviors impact them~\cite{datta15pets}.
Our work differs by looking at the accuracy of Google's stated inferences on real users.

Balebako et al.\ studied the effectiveness of PETs by examining how personalized the ads shown to browsers with PETs were compared to those shown to browsers without PETs installed~\cite{balebako12w2sp}.
In addition to differing by looking at real users, our work differs by looking at the Ad Settings interface instead of ads.

Small-scale anecdotal examinations of the accuracy of Ad Settings have appeared in the popular press~\cite{slate,dailydot}, as has a survey looking at the accuracy of Google's geo-location abilities~\cite{sel}.

\section{Methods}

With IRB approval, we conducted a survey that consisted of three types of information collection.  
First, we provided participants standard questions to which they responded.  We asked questions about their gender, age, browser usage, 
PETs usage, and opinions on the importance of privacy.

Second, we collected the inferred demographics displayed by Google's Ad Settings to compare to the participant's self-reported gender and age.  
We showed participants a screen shot of what this page looks like and included instructions on how to copy and paste the main content of the page. We asked them to paste it into a web form.  We used scripts to extract various variables from this page: inferred age from data from Google's services, inferred gender from Google's services, inferred age from data from across the web (Google's ad network), inferred gender from across the web, whether the user was logged into Google, and whether the user opted out of Google's interest-based ads.

Third, we conducted measurements of the participants' web browsers. We used an invisible iframe to have their browsers visit our server, which ran a series of tests to determine whether or not first party or third party cookies were blocked, whether Google Analytics cookies were blocked, and whether their browser was transmitting the DNT header.

Our survey yielded four measures of accuracy along with numerous factors they could be associated with, making a myriad of comparisons possible. %
To compensate for the multiple testing problem, we split our survey responses into an exploratory set and a confirmation set.
We used the exploratory set to identify associations that appear statistically significant under the $\chi^2$ test without adjusting for multiple tests.
We then tested just these associations on the confirmation data while adjusting for the total number of confirmation tests (nine) using the $\chi^2$ test with a Bonferroni correction.
For reasons of space, we report frequencies and accuracies over the whole data set despite comments on statistical significance referring to the two subsets of data.

\section{Results}

We recruited for our survey using Mechanical Turk with an offer to pay 500 participants.  
On Oct.\ 29, 2014, 
558 Turkers started our survey with 501 completing it.
We eliminated \py{len(all_responses_dirty) - len(all_responses_copy)}
responses for not correctly providing us with a copy of their Ad Settings page
and an additional \py{len(all_responses_copy) - len(all_responses_self_report)}
for not self-reporting gender and age.

Of the remaining responses, \py{len(all_responses_self_report) - len(all_responses)} of them were created using browsers with cookies disabled
for which Google displayed a message saying as much and that it had no inferences for the person.  
To simplify the presentation, we eliminated these responses as well, although they could be considered additional cases of Google opting to not make an inference.

We took the first \py{len(explore_responses)} of the remaining \py{len(all_responses)} responses to be our exploratory data set and the remaining \py{len(confirm_responses)}
responses to be our confirmation data set.
We use a temporal split of the data set to emphasize predictive ability.

\minisec{Account Settings}

Table~\ref{tbl:opt-out} shows how many of the remaining respondents opted out of tracking on Google services or Google ads across the web.
\begin{table}
\small
\centering
\py{
attributes_tabular(all_responses, 
                   [('copy: opt out: Ads on Google', 'Google'), 
                    ('copy: opt out: Google ads across the web', 'Across the web'),],
                   transpose=True)
}
\caption{Number of respondents who opted out or in of various forms of tracking by Google}
\label{tbl:opt-out}
\end{table}
Additionally, we found that \py{len(filter(lambda r: r['copy: opt out: YouTube'] == 'opted in', all_responses))} 
of the respondents who opted out of Google ads on services also opted in for getting ads on YouTube, presumably overriding the more general opt out for that service.

We found 
\py{len(filter(lambda r: r['copy: account status'] == 'logged in', all_responses))}
respondents to be logged into a Google account
and
\py{len(filter(lambda r: r['copy: account status'] == 'logged out', all_responses))}
to not be.
Since logged in users' profiles are available, which makes inferences easier, we break down all further results along the lines of logged in and out users.

\minisec{Overall Accuracy}

Table~\ref{tbl:sexes-age} shows both the self-reported and inferred genders and ages.
The values of ``Unknown'' and ``N/A'' are ones that Google lists, not comments from the authors on what we know or applicability.  
(``n/a'' is ours.)
Google gave one user an age range that overlapped with two other age ranges.  
We drop this range from further analysis.

\begin{table}
\small
\centering
\py{
latex_tabular.tabular(gender_counts + age_counts, header=attr_header, column_aligns="l"+"r"*(len(age_counts[0])-1))
}
\caption{Number of respondents with each value broken by being logged in or out}
\label{tbl:sexes-age}
\end{table}

Table~\ref{tbl:google-results} summarizes how often Google correctly stated the participants' sexes and ages.
We report the percentage of participants that Google got right, wrong, and skipped (by listing unknown or N/A).
The results show that Google skipped 100\% of participants who were logged out for Google services.
In these cases, Google got 0\% right, meaning Google is very
inaccurate in one sense, but, in different sense, Google's accuracy
cannot even be evaluated since Google did not try to make inferences
in this cases.
Google also skipped over 70\% of logged out participants for across the web.
The results also show that Google rarely made a wrong prediction.
From this, we conclude that Google is conservative in making predictions, but typically
right when it does so.

\begin{table}
\small
\centering
\py{tabular_corrects(all_in_responses, all_out_responses)}
\caption{Google's accuracy shown as the percentages Google got right, wrong, or skipped.  We treat as skipped those Google called ``Unknown'' or ``N/A''.}
\label{tbl:google-results}
\end{table}

\minisec{Demographics}

We checked whether the percentage Google got right was associated with actual gender or age of participants.
Table~\ref{tbl:demo-right} shows the results.
Note that a column of all 0s reflects that Google did not attempt to
make an inference for that combinations of factors (see
Table~\ref{tbl:google-results}), not Google guessing consistently incorrectly.
 
The results show that Google was right roughly as often for
females as males, with the difference exceeding 10\%
for inferences about age for logged out users across the web, where
the percentage Google got right was 17\% for females and just 6\% for
males, for an 11\% difference.
Given that Google only attempted to draw an age inference for 21\%
participants logged out for across the web, this difference might be
just noise from the small number of attempts.

For age, the largest drop in the number right from the overall number
is age for Google services for people who self-report an age of 65 or
more, which could be just noise  
given the small number of participants in that age bracket.
Focusing on the age brackets for which we have at least 50
participants (which cover ages from 18 to 44), we find the largest
differences to be between the age brackets to be for logged out
age across the web (25\% vs.\ 6\% right).
The results include two statistically significant associations for
age in the exploratory data set; neither of these differences proved
significant in our confirmation data set.

\begin{table}
\small
\py{
groups_prc_for_each([all_in_responses, all_out_responses], 
                    rights, 'correct', 
                    prepend_functional(demo_preds, ('payment', 'All', always(True))), True, 
                    prc_header,  p_values=True, 
                    explore_responses=[explore_in_responses, explore_out_responses], 
                    confirm_responses=[confirm_in_responses, confirm_out_responses])
}
\caption{The percentage that Google got right for each reported gender and age range.  Underlining shows associations with statistical significance in our exploratory data set.}
\label{tbl:demo-right}
\end{table}

\minisec{Computer Usage}

Table~\ref{tbl:usage-count-accuracy} shows the number of respondents with various usage conditions on the computer used to take our survey and the percentage of them about whom Google made correct inferences.
Some of these activities intuitively makes it more difficult to correctly make inferences about any one user of the computer since they imply that the computer has multiple users, which could pollute a model of any one of them.
The decrease in accuracy is sizable in some cases and reaches statistical significance in our exploratory data set for three conditions involving the clearing cookies.
However, none have a statistically significant association with Google's error rate in our confirmation data set.

\begin{table*}
\small
\centering
\begin{tabular}{@{}l rrrr c rrrrrrrrr@{}}
\toprule
  & \multicolumn{4}{c}{Counts} &\mbox{}\hspace{1em}\mbox{}& \multicolumn{8}{c}{Percent right}\\
\cmidrule(lr){2-5} \cmidrule(l){7-14}
  & \multicolumn{2}{c}{Yes} & \multicolumn{2}{c}{No} && \multicolumn{2}{c}{Sex services} & \multicolumn{2}{c}{Sex web} & \multicolumn{2}{c}{Age services} & \multicolumn{2}{c}{Age web}\\ & in & out & in & out && in & out & in & out & in & out & in & out\\
\midrule
All (baseline)                  & n/a & n/a & n/a & n/a&& 66 & 0 & 74 & 21 & 67 & 0 & 65 & 11\\
Shared computer                 & 62 & 19 & 335 & 65   && 63 & 0 & 77 & 21 & 71 & 0 & 69 & 11\\
Shared account                  & 28 & 12 & 369 & 72   && 54 & 0 & 75 & 25 & 64 & 0 & 61 & 17\\
Other users in a week           & 107 & 24 & 290 & 60  && 62 & 0 & 74 & 21 & 69 & 0 & 69 & 17\\
Other users yesterday           & 61 & 17 & 336 & 67   && 59 & 0 & 67 & 18 & 66 & 0 & 64 & 18\\
Cleared cookies today/yesterday\mbox{}\hspace{1em}\mbox{} & 58 & 36 & 339 & 48   && 64 & 0 & \underline{66} & 8 & 64 & 0 & 59 & 3\\
Cleared cookies on close        & 33 & 17 & 364 & 67   && 61 & 0 & 64 & 18 & \underline{55} & 0 & \underline{48} & 6\\
Private mode                    & 18 & 6 & 379 & 78    && 56 & 0 & 61 & 0 & 61 & 0 & 56 & 0\\
\bottomrule
\end{tabular}
\caption{The number of respondents with each computer usage characterization and the percentage that Google got right for just respondents with each computer usage characterization.
For the percentages, underlining shows associations with statistical significance in our exploratory data set (i.e., a statistically significant difference from the baseline).}
\label{tbl:usage-count-accuracy}
\end{table*}

\minisec{Attitudes}

Table~\ref{tbl:attitude} in the appendix shows the associations between the respondents' attitudes toward tracking and Google's accuracy.
Our exploratory analysis found no significant associations.

\minisec{PETs}

Table~\ref{tbl:pets-joint} shows the usage of various PETs and the number Google got right for users of each PET.
Above the bar are the self-reported usage habits of PETs by respon\-dents.
Below the bar are our server's measurements.
For these measurements, ``empty'' means that our server did not detect a visit from the respondent (e.g., due to network loss).
Unfortunately, the small number of users of some of the PETs limits our abilities to draw conclusions about them.

\begin{table*}
\footnotesize
\centering
\begin{tabular}{@{}l rrrrrrrr c rrrrrrrr@{}}
\toprule
  & \multicolumn{8}{c}{Counts} &\mbox{}\hspace{1em}\mbox{}& \multicolumn{8}{c}{Percent right}\\
\cmidrule(lr){2-9} \cmidrule(l){11-18}
 & \multicolumn{2}{c}{Yes} & \multicolumn{2}{c}{No} & \multicolumn{2}{c}{I don't know} & \multicolumn{2}{c}{empty}    && \multicolumn{2}{c}{Sex services} & \multicolumn{2}{c}{Sex web} & \multicolumn{2}{c}{Age services} & \multicolumn{2}{c}{Age web}\\ 
& in & out & in & out & in & out & in & out && in & out & in & out & in & out & in & out\\
\midrule
All (baseline) &n/a&n/a&n/a&n/a&n/a& n/a&n/a& n/a &&                   66 & 0 & 74 & 21 & 67 & 0 & 65 & 11\\                       
AdBlock & 216 & 45 & 138 & 28 & 40 & 11 & 3 & 0   &&                  66 & 0 & \underline{67} & \underline{7} & 65 & 0 & \underline{56} & 4\\     
Ghostery & 18 & 8 & 333 & 69 & 45 & 7 & 1 & 0   &&                    56 & 0 & 33 & 0 & 61 & 0 & 28 & 0\\                               
NoScript & 12 & 10 & 317 & 54 & 66 & 19 & 2 & 1   &&                  42 & 0 & 75 & 0 & 58 & 0 & 58 & 0\\                               
DoubleClick opt out & 15 & 4 & 298 & 66 & 84 & 14 & 0 & 0   &&        53 & 0 & 60 & 50 & 47 & 0 & 60 & 25\\                  
Webpage opt out & 82 & 20 & 246 & 47 & 69 & 16 & 0 & 1   &&           65 & 0 & 61 & 30 & 60 & 0 & \underline{57} & 15\\             
DNT set & 100 & 34 & 217 & 33 & 80 & 17 & 0 & 0   &&                  57 & 0 & 68 & 18 & 65 & 0 & 65 & 6\\                               
\midrule                                                                                                                               
DNT sent & 53 & 22 & 339 & 55 & n/a & n/a & 5 & 7   &&                    57 & 0 & 68 & 23 & 62 & 0 & 57 & 9\\                              
1st cookies off & 22 & 11 & 370 & 66 & n/a & n/a & 5 & 7   &&             45 & 0 & 68 & 36 & 68 & 0 & 68 & 9\\                       
3rd cookies off & 24 & 13 & 367 & 64 & n/a & n/a & 6 & 7   &&             50 & 0 & 71 & 38 & 71 & 0 & 75 & 15\\                      
Google cookies off & 39 & 17 & 353 & 60 & n/a & n/a & 5 & 7   &&          54 & 0 & 64 & 24 & 59 & 0 & 59 & 6\\                    
\bottomrule
\end{tabular}
\caption{Number of users of each PET and the percentage that Google got right for users of each PET.  Underlining shows associations with statistical significance in our exploratory data set.  The percentages correct are broken by PET showing the percentage correct for just those participants who answered ``yes'' to having the PET or for whom we detected the PET.}
\label{tbl:pets-joint}
\end{table*}

Looking at AdBlock, in our exploratory data set, we found a statistically significant reduction in the accuracy of Google for data from across the web for sex both when logged in and out.
A significant reduction in the accuracy for age across the web also exists, but only when logged in.
The largest of these, for logged out sex across the web, was from 21\% down to 7\%, a drop of 14\%.
The only other PET to get statistical significance is using webpage opt outs, and only in the case of age across the web when logged in.
None of these differences proved statistically significant in our confirmation data set.

For two PETs, Ghostery and NoScript, the percentage that Google got
correct is always 0\% for logged out users.
This is the best a PET can do in that logged in users may provide
their demographics to Google directly, circumventing the PET.
We cannot, from our observational data, conclude that these
PETs caused the decrease.

\section{Discussion}

We have no way of knowing whether the inferences shown on Ad Settings are the same as those actually used by Google for ad targeting, and prior work suggests that Ad Settings does not provide information about how Google remarkets to users based upon prior webpage visits~\cite{datta15pets} (a limitation made explicit on the Ad Settings page after the publication of~\cite{datta15pets}).
Nevertheless, we find it noteworthy that Google rarely shows inferences for logged out users.
We can only conjecture as to the reason, but perhaps one's web browsing behavior is not as visible to or interpretable by Google as some fear.
Unfortunately, since conducting our survey, Google has disabled Ad Settings for logged out users, precluding the possibility of further studying this phenomenon.

We relied upon self-reports of age and gender for ground truth, of PETs usage, and of behavior while looking for factors associated with Google's accuracy.
Self-reports of PETs usage, in particular, may be inaccurate due to the obscurity of PETs and the possibility that a shared browser may use one without the respondent's knowledge.  
Furthermore, our server's attempts to detect PETs usage by examining the behavior of respondents' browsers could have measurement errors from factors such as network loss.

Mechanical Turkers might not be representative of standard web users.  In particular, they may be more likely to use PETs or security measures due to the heavy use of their browsers for Turking.
Furthermore, they may visit an atypically large number of webpages unassociated with their demographics to fulfill their Turking tasks.

Our exploratory results suggest that cookie clearing and AdBlock %
may be associated with decreasing Google's accuracy. %
Using observational data, we cannot claim that they cause the decrease.

Future work includes 
running experiments to determine whether PET usage is the cause of such decreases in accuracy
and
conducting a larger-scale observational studies to bring larger number of PET users and cookie clearers into the sample.  
We hope this will allow us to find statistically significant associations, which in some cases appear unobtainable due to the small number of respondents with privacy-seeking behaviors (Tables~\ref{tbl:usage-count-accuracy} and~\ref{tbl:pets-joint}).

\section*{Acknowledgements}

This research was supported by the U.S.\ National Science Foundation
(NSF) grants CNS 1065240 and CNS 1514509.  The views and conclusions
contained in this document are those of the authors and should not be
interpreted as representing the official policies, either expressed or
implied, of any sponsoring institution, the U.S.\ government or any
other entity.

\bibliographystyle{acm}
\bibliography{local}

\appendix

\section{Additional Details}

For the purposes of Table~\ref{tbl:usage-count-accuracy}, we defined a ``shared computer'' to be one that respondent described as ``Regularly used by multiple workers at a place of employment'', ``Regularly used by multiple members of a family'', or ``Regularly used by many people in a public place (library, Internet cafe, etc.)'', but not as ``Regularly used only by me'' nor as ``None of the above''.

For the purposes of Table~\ref{tbl:attitude}, we defined ``Concerned about tracking'' as answering with a 4 or 5 (very concerned) on a 5-point scale to the question ``How concerned are you about online tracking of your behavior?''
We defined ``Confidence about avoiding it'' as a 4 or 5 (very confident) to the question ``If you have taken steps to prevent online tracking of your behavior, how confident are you that it prevents online tracking?''

\section{Additional Data}

\begin{figure*}
\begin{center}
\includegraphics[width=\textwidth]{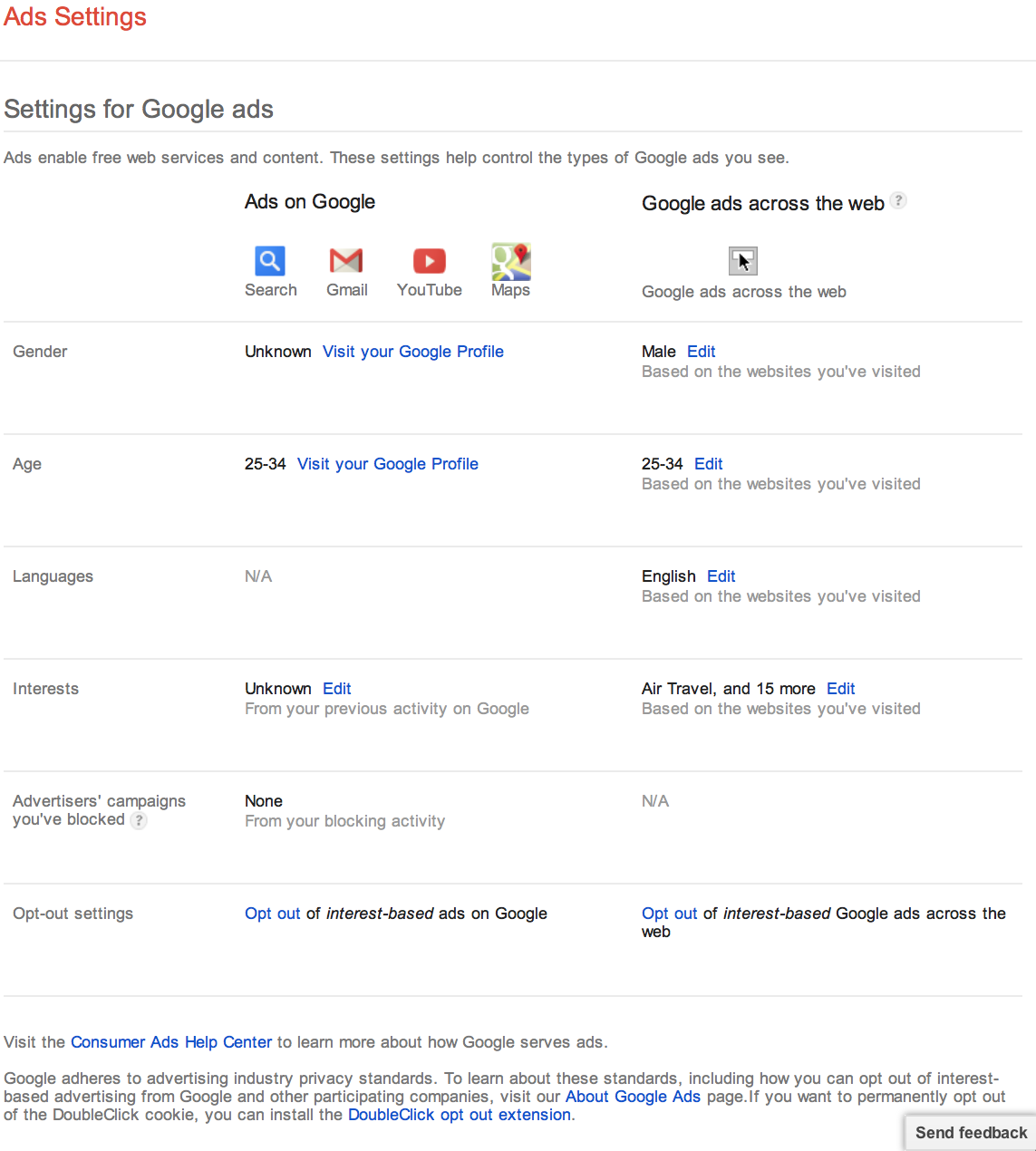}%
\end{center}
\caption{Screen shot of Google's Ad Settings webpage while logged in with a Google account using Safari.  \textcopyright Google}
\label{fig:settings-saf-tmc}
\end{figure*}

The following tables show the responses we received to various questions on our survey.
Figure~\ref{fig:settings-saf-tmc} provides a larger screen shot of Google's Ad Settings.

\begin{table}[!hb]
\py{
tabular_corrects(filter(lambda r: r['What is your gender?'] == 'Female',
                        all_in_responses), 
                 filter(lambda r: r['What is your gender?'] == 'Female',
                        all_out_responses))
}
\caption{Google's accuracy shown in percentages for females.}
\end{table}

\begin{table}[!hb]
\py{
tabular_corrects(filter(lambda r: r['What is your gender?'] == 'Male',
                        all_in_responses), 
                 filter(lambda r: r['What is your gender?'] == 'Male',
                        all_out_responses))
}
\caption{Google's accuracy shown in percentages for males.}
\end{table}

\py{
dump(['Did you use this computer yesterday?', 
      'In the past week, on how many days did you use this computer?', 
      'Which best describes this computer?', 
      'In the past week, on how many days did someone other than you use this computer?', 
      'If anyone else used the computer you are currently using within the last week, did that person(s) use a different user account from the one you are currently using?', 
      'Did anyone else use this computer yesterday?',])
}

\py{dump(remaining)}

\begin{table*}
\centering
\py{
groups_prc_for_each([all_in_responses, all_out_responses], 
                    rights, 'correct', 
                    prepend_functional(attitude_preds, ('payment', 'All (baseline)', always(True))), True, 
                    prc_header,  p_values=True, 
                    explore_responses=[explore_in_responses, explore_out_responses], 
                    confirm_responses=[confirm_in_responses, confirm_out_responses])
}
\caption{The percentage that Google got right for respondents with each attitude}  
\label{tbl:attitude}
\end{table*}

\end{document}